\documentclass[final,1p,times]{elsarticle}
\usepackage{amssymb}
\usepackage{amsmath}
\usepackage{multirow}
\usepackage{siunitx}
\usepackage{floatrow}
\usepackage{upgreek}
\usepackage{subfig}
\usepackage{caption}
\usepackage{floatrow}
\usepackage{placeins}   
\usepackage{xcolor}
\usepackage{xcolor}

\usepackage[modulo, pagewise]{lineno}

\journal{elsevier}

\usepackage{soul}
\begin{document}

\begin{frontmatter}



% \title{Solute strengthening in discrete dislocation dynamics simulations of fcc high entropy alloys}
\title{The effect of local chemical ordering on dislocation activity in multi-principle element alloys: a three-dimensional discrete dislocation dynamics study}


\author[JHU]{Markus Sudmanns\corref{cor1}}
\ead{msudmanns@jhu.edu}
\author[JHU]{Jaafar A. El-Awady}
\ead{jelawady@jhu.edu}

\cortext[cor1]{Corresponding author: }

\address[JHU]{Department of Mechanical Engineering, The Whiting School of Engineering, The Johns Hopkins University, Baltimore, MD 21218, USA}

\begin{abstract}

The exceptional combination of strength and ductility in multi-component alloys is often attributed to the interaction of dislocations with the various solute atoms in the alloy. To study these effects on the mechanical properties of such alloys there is a need to develop a modeling framework capable of quantifying the effect of these solutes on the evolution of dislocation networks. Large scale three-dimensional (3D) Discrete dislocation dynamics (DDD) simulations can provide access to such studies but to date  
no relevant approaches are available that aim for a complete representation of real alloys with arbitrary chemical compositions.
Here, we introduce a formulation of dislocation interaction with substitutional solute atoms in fcc alloys in 3D DDD simulations that accounts for solute strengthening induced by atomic misfit as well as fluctuations in the cross-slip activation energy.
Using this model, we show that local fluctuations in the chemical composition of various CrFeCoNi-based multi-principal element alloys (MPEA) lead to sluggish dislocation motion, frequent cross-slip and alignment of dislocations with solute aggregation features, explaining experimental observations related to mechanical behavior and dislocation activity.
It is also demonstrated, that this behavior observed for certain MPEAs cannot be reproduced by assuming a perfect solid solution.
The developed method also provides a basis for further investigations of dislocation plasticity in any real arbitrary fcc alloy with substitutional solutes.

\end{abstract}


\end{frontmatter}

%\linenumbers


\section{Introduction}
\label{1_Intro}
The interaction of dislocations with solute atoms in alloys is one of the major contribution to material strengthening.
One of the reasons for this strengthening effect is attributed to the atomic misfit of the different alloy components, which introduce local lattice distortions increasing the stress required to move a dislocation \cite{varvenne_theory_2016, leyson_quantitative_2010}.
Local spacial variation in the chemical composition of the alloy are commonly observed in multi-principal element alloys (MPEAs) after heat treatment \cite{Zhang2020, ma2020unusual, ma2019tailoring} and can also emerge due to manufacturing processes, e.g. in additive manufacturing of MPEAs, stainless steels or nickel alloys \cite{wang_additively_2018,liu_dislocation_2018,ZHANG2018200,e20120937} .
MPEAs contain a nominally near-equiatomic chemical composition of three or more alloying elements and have been show to exhibit an unprecedented combination of favorable material properties, especially strength and ductility (cf. \cite{ma2019tailoring,zhang2014microstructures,MIRACLE2017448}).
Spacial variations in the chemical composition have been commonly observed experimentally in MPEAs and are described as local chemical ordering (LCO) or local short range order (SRO) for the short and medium ranges and is said to lead to the increase in the yield strength and hardening due to size and modulus mismatches \cite{Zhang2020, li2019strengthening, ding_tuning_2019}.
Depending on the specific alloy composition, LCO can promote complex dislocation dynamics characterized by sluggish dislocation glide and complex dislocation interactions \cite{ma2020unusual, shim_nanoscale_2019}. Furthermore, for some face-centered cubic (FCC) MPEAs intense cross-slip was observed, which was explained by the resistance against dislocation glide on primary slip systems caused by sluggish dislocation motion \cite{ding_tuning_2019, ding2019real}. However, others showed a predominately planar glide \cite{Zhang2020}. 

The mobility of individual dislocations in a MPEA solute field was recently studied by molecular dynamics (MD) simulations, which showed a wavy and sluggish dislocation motion caused by LCO  \cite{li2019strengthening, Antillon2020d, li2020core}. Additionally, MD simulations of cross-slip behavior in arbitrary FCC alloys showed a significant fluctuation in the activation energy of cross-slip in random alloys \cite{nohring_dislocation_2017}.
Lower activation barriers dominate the overall cross-slip activity for moderate applied stresses, suggesting that the stacking fault energy alone is not a sufficient measure to evaluate cross-slip activity in FCC alloys \cite{nohring_cross-slip_2018}.
However, MD simulations are limited in time and length scale which prohibits a full understanding of the evolution of ensembles of dislocations in dimensions relevant to experimental observations.

To be able to develop a reliable prediction of the mechanical properties of MPEAs, all previously mentioned aspects have to be combined into a comprehensible explanation for the relationship between the dislocation microstructure and the mechanical properties.
Three-dimensional (3D) Discrete Dislocation Dynamics (DDD) simulations can provide an understanding of the influence of solidification induced solute segregation on the evolution of dislocation ensembles and their interplay towards mechanical properties. 
However, most investigations in DDD have so far been limited to pure metals since no relevant approach towards explicitly considering the effect of solid solution atoms on the evolution of the dislocation microstructure has been made. 
Therefore, micromechanical simulations of large dislocation ensembles including a modeling of dislocation solute interaction in real alloys have not been achieved to date.
Approaches which go beyond pure metal in DDD include interstitial atoms such as hydrogen, e.g. presented in \cite{gu_quantifying_2018, gu_theoretical_2020}, dislocation-precipitate interaction as in \cite{shin2003dislocation,lehtinen2016multiscale}, or investigate the motion of a single dislocation under stochastic forces \cite{zhai2019properties}. 
In a phase-field dislocation dynamics model the operation of a Frank-Read (FR) source in a body-centered cubic (BCC) MPEA has been investigated, showing a significant influence of solute composition fluctuations on the FR source operation \cite{smith2020effect}.

To address these limitations, here, the mechanical behavior and dislocation activity in MPEAs incorporating fluctuations in chemical composition are analyzed with a new 3D DDD framework that accounts explicitly for dislocation-solid solution atom interactions as well as effects on cross-slip activation.
Using an in-house modification of the open source code ParaDiS \cite{arsenlis_enabling_2007}, we incorporate recent approaches towards accounting for solute strengthening on dislocation plasticity in FCC alloys \cite{varvenne_theory_2016}.
Furthermore, we account for observations of fluctuations in the cross-slip activation energy reported from MD simulations \cite{nohring_dislocation_2017}.
Through this new approach,  the local solute composition variation in CrFeCoNi-based MPEAs observed experimentally are investigated in order to predict the mechanisms governing plastic deformation and resulting mechanical response in bulk DDD simulations and provide an in depth explanation of  experimental observations \cite{li2019strengthening, ding_tuning_2019, shim_nanoscale_2019}.

\section{Computational Method}
\label{2_Meth}
All simulations here are conducted using an in-house version of the 3D DDD open-source code ParaDiS \cite{arsenlis_enabling_2007}. The open source code was significantly modified to avoid any artificial non-planar dislocation glide or collisions events in FCC crystals as well as to incorporate atomistically-informed cross-slip mechanisms as described in \cite{hussein_microstructurally_2015}. To model dislocation plasticity in FCC MPEAs, three additional aspects must also be accounted for within the framework of 3D DDD simulations. First, the stress induced by alloying atoms on the dislocations as they glide in the FCC lattice (i.e. dislocation mobility). Second, the effect of local fluctuations in the chemical composition on the cross-slip activation barrier \cite{nohring_dislocation_2017, nohring_cross-slip_2018}. Third, the effect of local chemical ordering (LCO). The theoretical treatment of these effects within the framework of 3D DDD simulations are discussed in the following subsections.

\subsection{Solute effects on dislocation mobility} \label{sec:Methods-Mobility}

To account for the solute effect on the dislocation critical resolved shear stress (CRSS) we utilize the theoretical model developed by Varvenne \textit{et al.} for a dislocation segment gliding in a single phase alloy with an arbitrary composition \cite{varvenne_theory_2016, varvenne_solute_2017}. In this model, the CRSS for a dislocation segment is derived from the interaction energies between the solutes and the dislocation segment. It is assumed that the solute-dislocation interactions arise primarily form the elastic interactions between solutes and the pressure field induced by the dislocation segment. On the other hand, misfit strains and effects related to changes in the elastic modulus are neglected in this model. Additionally, the model considers that the solute-dislocation interactions are independent of the dislocation orientation \cite{varvenne_solute_2017}. This is justified by the argument that even non-edge dislocations in FCC metals dissociate into two partials, each having an edge component, which induces a pressure field in the crystal lattice \cite{ma_computationally_2015, yasi_first-principles_2010}. Finally, the model also neglects dislocation core interactions with the solutes, including diffusion processes at time scales similar to that of the dislocation motion.

In this model, the interaction between solutes and a dislocation segment can be quantified by an energy barrier $\Delta E_b$, which the dislocation must overcome after reaching the related zero temperature strength $\tau_\mathrm{c}^0$ (i.e. dislocation Peierls stress) \cite{varvenne_theory_2016}.  This process is assisted by thermal activation and the CRSS of the dislocation can be evaluated as follows \cite{LEYSON20123873}:
\begin{equation}
    \tau_\mathrm{c}(T,\dot{\varepsilon}) = \tau_\mathrm{c}^0\text{exp}\left(-\frac{1}{0.51}\frac{kT}{\Delta E_b}\text{ln}\frac{\dot{\varepsilon}_0}{\dot{\varepsilon}}\right)
    \label{eq:solute_stress}
\end{equation}
where $T$ is the temperature, $k$ is Boltzmann's constant, $\dot{\varepsilon}$ is the experimental strain rate, and $\dot{\varepsilon}_0 = 10^{4}\text{s}^{-1}$ is a reference strain rate \cite{varvenne_theory_2016}. Additionally, the energy barrier and the dislocation Peierls stress can be expressed as \cite{varvenne_theory_2016}:
\begin{equation}
\tau_\mathrm{c}^0 = 0.051\alpha^{-\frac{1}{3}}\mu\left(\frac{1+\nu}{1-\nu}\right)^{\frac{4}{3}}f_1(w_c)\times\left[\frac{\sum_i c_i\left(\Delta\bar{V}^2_i+\sigma^2_{\Delta V_i}\right)}{b^6}\right]^{\frac{2}{3}}
\label{eq:tauzero}
\end{equation}
and
\begin{equation}
\Delta E_b = 0.274\alpha^{\frac{1}{3}}\mu\left(\frac{1+\nu}{1-\nu}\right)^{\frac{2}{3}}f_2(w_c)\times\left[\frac{\sum_i c_i\left(\Delta\bar{V}^2_i+\sigma^2_{\Delta V_i}\right)}{b^6}\right]^{\frac{1}{3}}
\label{eq:deltaeb}
\end{equation}
where $\mu$ is the shear modulus, $\nu$ the Poisson ratio, $b$ is the Burgers vector magnitude, $f_1(w_c) = 0.35$ and $f_2(w_c) = 5.70$ are core coefficients \cite{varvenne_theory_2016}, and $\alpha=0.125$ is the line-tension parameter for various FCC MPEAs \cite{Yin2020b}. 
Using Eqns.\,({\ref{eq:solute_stress}})-({\ref{eq:deltaeb}}) with these parameters, the yield stress estimated analytically was shown to be in good agreement with experimental measurements for various FCC alloys at different temperatures \cite{varvenne_theory_2016,varvenne_solute_2017,jiao2018thermo}.
The term $\sum_i c_i\left(\Delta\bar{V}^2_i+\sigma^2_{\Delta V_i}\right)$ in Eqs. (\ref{eq:tauzero}) and (\ref{eq:deltaeb}) quantifies the misfit of solute $i$ compared to the average alloy, where $\Delta\bar{V}_i = V_i - \bar{V}$ is the misfit volume of each solute, $V_i$ is the atomic volume of solute $i$ calculated from the atomic radii, $\bar{V} = \sum_i c_i V_i$ is the average volume, $c_i$ the solute concentration, and $\sigma_{\Delta V_i}$ is the standard deviation of the misfit volume \cite{varvenne_theory_2016, varvenne_solute_2017}. The standard deviation of the misfit volume represents the fluctuations in the local chemical environment around the solute site that affect the solute-dislocation interactions in non-dilute alloys such as the ones studied here. The atomic radii of the different elements used in the current study are listed in Table\,\ref{tab:atomic_radii}.

\begin{table}[h]
\centering
\caption{The atomic radii, $r_i$, from \cite{senkov_effect_2001} and the atomic volume, $V_i = a_i^3/4$, where the lattice constant $a_i = (4r_i)/\sqrt{2}$, for the different elements used in the current study.}
\begin{tabular}{@{\extracolsep{4pt}}lll@{}}
\hline
Element & Atomic radius (nm) & Atomic volume (nm$^3$)\\
\hline
Co & 0.12510 & 0.0110751\\
Pd & 0.13754 & 0.0147184\\
Mn & 0.135 & 0.013918\\
Fe & 0.12412 & 0.0108168\\
Ni & 0.12459 & 0.0109402\\
Cr & 0.12491 & 0.0110247\\
Cu &  0.12510 & 0.0118078\\
\hline
\end{tabular}
\label{tab:atomic_radii}
\end{table}

It should be noted that Eq. (\ref{eq:solute_stress}) is valid when the applied shear stress $0.2 < \tau/\tau_\mathrm{c}^0 < 1$, i.e. for intermediate temperatures between $\approx\,$300K and $\approx\,$1000K. For higher temperatures or lower stresses (i.e. $\tau_\mathrm{c}/\tau_\mathrm{c}^0 < 0.2$), Leyson and Curtin suggested that a power-law formulation can better capture the experimental results \cite{LEYSON20123873}. However, since the current study is restricted to room temperature response, only the formulation for intermediate temperatures is considered here. 

In the framework of 3D DDD simulations the solute concentration dependent dislocation segment CRSS as given by Eq. (\ref{eq:solute_stress}) can be converted to a solute force on each dislocation node according to $\mathbf{f}_\text{sol} = - \tau_\mathrm{c}(T,\dot{\varepsilon})\,l\,b\,\mathbf{f}_\text{l}/\|\mathbf{f}_\text{l}\|$, where the dislocation nodal forces $\mathbf{f}$ are projected on the glide plane (i.e. $\mathbf{f}_\textbf{n} = \mathbf{n}\times\left(\mathbf{f}\times\mathbf{n}\right)$) then projected in the direction normal to the line direction of the dislocation segment (i.e. $\mathbf{f}_\textbf{l} = \mathbf{l}\times\left(\mathbf{f}_\textbf{n}\times\mathbf{l}\right)$). Here, $\mathbf{n}$ is the dislocation segment slip plane normal vector, $\mathbf{l}$ is the dislocation segment line direction unit normal vector, and $l$ its length. Every simulation time-step the magnitude of the solute force $\|\mathbf{f}_\text{sol}\|$ is compared with the magnitude of the projected nodal force $\|\mathbf{f}_\text{l}\|$. If $\|\mathbf{f}_\text{sol}\| < \|\mathbf{f}_\textbf{l}\|$, then the effective nodal force is calculated as $\mathbf{f}_\text{eff} = \mathbf{f} - \mathbf{f}_\text{sol}$. Otherwise, if $\|\mathbf{f}_\textbf{l}\| \leq \|\mathbf{f}_\text{sol}\|$ then the effective nodal force is calculated as $\mathbf{f}_\text{red} = \mathbf{f} - \mathbf{f}_\textbf{l}$, which practically means the dislocation node will not move in the glide direction during this time step. Note that the nodal force $\mathbf{f}$ is not set to zero in order to allow the dislocation node to move in the dislocation line direction, which is necessary to maintain consistent topological operations each simulation time-step (e.g. dislocation node collisions during junction formation or annihilation).

The effective nodal force $\mathbf{f}_\text{eff}$ is related to the nodal velocity $\mathbf{v}$ by a dislocation mobility law in the form of $\mathbf{v} = M(\mathbf{f}_\text{eff})$, where $M$ is a mobility function whose arguments contain the effective nodal force as well as a dislocation drag coefficient $B$.
For simplicity, the simulations shown in this study incorporate a dislocation mobility law for pure Ni. 
The mobility law reproduces a viscous drag mechanism, which is characterized by a linear relation between nodal force and nodal velocity.
To our knowledge, composition-dependent dislocation mobility laws for the MPEAs under consideration here are not yet available in the literature to date.
However, the conclusions of the work do not change when using a concentration-dependent dislocation mobility law for FeNiCr-based stainless steels obtained by molecular dynamics simulations, which confirm the linear dislocation mobility in such alloys \cite{Chu2020} (see \ref{sec:appendix} for more details).

\subsection{Cross-slip in fcc MPEAs}\label{sec:Methods-CrossSlip}

Atomistic simulations of FCC solid solution alloys have shown that the cross-slip activation energy, $\Delta E_\text{act}$, is significantly lower than estimates based on the average elastic and stacking fault properties of the alloy, and is strongly influenced by local fluctuations in the spatial arrangement of the solutes. This leads to fluctuations in $\Delta E_\text{act}$ about an average activation energy barrier $\Delta E_\text{act, avg}$ \cite{nohring_dislocation_2017, nohring_cross-slip_2018}. 

In the current 3D DDD framework, the cross-slip frequency is calculated through an Arrhenius-like relationship according to \cite{hussein_microstructurally_2015}:
\begin{equation}
     f = \omega_a\frac{L}{L_0}\mathrm{exp\left({-\frac{\Delta E_\text{act}}{k_bT}}\right)},
     \label{eq:cs_frequency}
 \end{equation}
where $L$ is the length of the dislocations within $\pm 15^\circ$ of the screw orientation, $\omega_a$ is the attempt frequency, $L_0 = 1\,\upmu$m is a reference length \cite{kubin1992dislocation}, and the cross-slip activation energy is:
\begin{equation}
    \Delta E_\text{act} = E_a-V_a\Delta\tau_\mathrm{E}.
\end{equation}
Here $E_a$ is the energy required to form a constriction on the screw dislocation and is dependent on the cross-slip type (i.e. bulk, repulsive or attractive intersection, or surface induced cross-slip \cite{hussein_microstructurally_2015}), $V_a$ is the activation volume, and $\Delta\tau_\mathrm{E}$ is the difference between the Escaig stress (the stress component that controls the distance between the two partial dislocations) on the primary and cross-slip planes.
Here, bulk-type cross-slip denotes the traditional probabilistic, thermally activated cross-slip mechanism, whereas intersection-type cross-slip represents a mechanism that is preferentially observed at screw dislocation intersections in FCC \cite{hussein_microstructurally_2015}. Surface induced cross-slip is not active in the current simulations, since none of them incorporate free surfaces.

There is still a lack of quantitative analysis in the literature on the correlation between specific solute clusters and the resulting decrease or increase of the cross-slip activation energy or the Escaig stress. This prevents the development of a physically-based model for solute concentration dependent cross-slip in 3D DDD simulations. Therefore, a simplified first approach is utilized in the current 3D DDD framework to account for the energy fluctuation reported by atomistic simulations \cite{nohring_dislocation_2017, nohring_cross-slip_2018}. In this approach the cross-slip activation energy, $\Delta E_\text{act}$, for bulk-type cross-slip is assumed to vary randomly within $\pm50\%$ of a mean value independent of the specific local solute concentration were the dislocation segment under consideration is, which is reasonably close to the range reported by atomistic simulations \cite{nohring_dislocation_2017}. 
For intersection-type and surface-type cross-slip, to our knowledge no studies exist on their activation energies in solid solution alloys. 
Furthermore, since the MD simulations in \cite{nohring_dislocation_2017, nohring_cross-slip_2018} focused on the traditional thermally activated cross-slip mechanism, we restrict the aforementioned approach to bulk-type cross-slip. 
The cross-slip parameters used in the current study are listed in Table\,\ref{tab:cross_slip_activation_energy}, which are values computed for pure Ni.
The reason for the use of pure metal cross-slip parameters is the lack of reported parameters in literature for the alloys under consideration here. 
However, the current model can be easily adapted for the use of more accurate parameters, should they become available.
\begin{table}[h]
\centering
\caption{The cross-slip parameters used in the current study.}
\begin{tabular}{@{\extracolsep{4pt}}lll@{}}
\hline
Cross-slip type & Energy barrier $E_a$ $[eV]$& Activation volume $V_a$\\
\hline
Bulk & 0.4 \cite{RAO2017188} & $20b^3$\\
Surface & 0.2 \cite{RAO20132500} & $20b^3$\\
Hirth-lock & 0.2 \cite{RAO20117135} & $20b^3$\\
Lomer-Cottrel Lock & 0.6 \cite{RAO20117135} & $20b^3$\\
Glide lock & 0.5 \cite{RAO20117135} & $20b^3$\\
\hline
\end{tabular}
\label{tab:cross_slip_activation_energy}
\end{table}

\subsection{Accounting for local chemical ordering in MPEAs}
\label{sec:solute_aggregation}

Experimental studies have shown the formation of local chemical ordering (LCO) on short and medium ranges in some MPEAs, e.g. in \cite{ding_tuning_2019,shim_nanoscale_2019,chen2021direct}. Examples of LCOs in CrFeCoNi-based MPEAs include nanoscale solute aggregation in CrFeCoNiPd and lamella-like nanoscale structures in CoCu$_{1.71}$FeMnNi. Accordingly, it is important to incorporate such LCO within the framework of 3D DDD to faithfully model dislocation plasticity in such MPEAs. In the following we discuss how both types of LCOs are modeled in the current 3D DDD simulations.

\subsubsection{Nanoscale solute clusters in MPEAs}
By replacing the Mn in the Cantor alloy CrFeCoNiMn with the larger Pd, thus generating equi-atomic CrFeCoNiPd, an increased aggregation of constituting elements up to a chemical concentration of 58\% was observed in experiments \cite{ding_tuning_2019}
Those chemical aggregations are confirmed by first principle calculations \cite{tran2021stability} and are described as ``concentration waves'' with a length scale on the order of nanometers \cite{ding_tuning_2019}.
The observed random local solute aggregation can be modeled in 3D DDD by discretizing the 3D volume into voxels and assigning each voxel a solute composition chosen randomly from a Dirichlet distribution corresponding to the experimentally reported variations in CrFeCoNiPd \cite{ding_tuning_2019}. This results in concentration variations between $5\%$ and $50\%$ from the uniform concentration of the alloy. The effect of the local chemical composition on the dislocation glide resistance is represented by the solute misfit volume $\Delta V_i$ as well as its standard deviation $\sigma_{\Delta V_i}$ for which we estimate $\sigma_{\Delta V_i} = \Delta\bar{V}_i / 3$ based on the variation of the fluctuations in solute concentration given in \cite{ding_tuning_2019}. Hereafter, simulations with this type of LCO are referred to as \textit{Configuration-I}. 

Due to limitations in system resolution, and thus computational cost, cluster sizes on the order of a few nanometers cannot be numerically captured in the current DDD simulations, which typically have an average dislocation segment size on the order of 10s of Burgers vectors. Thus, for practical reasons, only clusters on the size of 10s of nm are considered here. 
In these simulations the fluctuations in the cross-slip activation energy barriers as discussed in Section \ref{sec:Methods-CrossSlip} are accounted for.

\subsubsection{Two-phase lamella structures in MPEAs}
The two-phase lamella structure observed experimentally in CoCu$_{1.71}$FeMnNi with a spacing of $\SI{100}{\nm}$ \cite{shim_nanoscale_2019} can also be modelled in 3D DDD simulations by introducing lamellae that are arranged parallel to the $[100]$ plane and assigning each lamella with the appropriate solute composition reported experimentally. 
Thus, the alloy is represented in our model by a 100\,nm spaced lamella structure, consisting of two separate solid solutions using chemical compositions reported in \cite{shim_nanoscale_2019} with no further underlying chemical substructure.
The two phases are characterized as a Co-Fe rich phase and a Cu rich phase. The estimate of a dislocation segment CRSS in each phase according to Eq.\,(\ref{eq:solute_stress}) is $\SI{118}{\MPa}$ for the Co-Fe rich phase and $\SI{91}{\MPa}$ for the Cu rich phase. 
The influence of strongly ordered nanoscale solute aggregations on the cross-slip activation energy has not been addressed from MD analysis (e.g. \cite{nohring_dislocation_2017, nohring_cross-slip_2018}). 
Additionally, to our knowledge no cross-slip activation energies have been reported in literature for either phase. 
Accordingly, as a first order assumption the cross-slip activation energy barrier reported in Table\,\ref{tab:cross_slip_activation_energy} were used for both phases in the current simulations. Hereafter, simulations with this type of LCO are referred to as \textit{Configuration-II}.

As comparison and reference cases, we also show simulations assuming a homogeneous chemical composition in CrFeCoNiPd and CrMnFeCoNi without any local variation in chemical composition and using the cross-slip activation energy barrier reported in Table\,\ref{tab:cross_slip_activation_energy} for both alloys.
Hereafter, simulations containing a homogeneous chemical composition will be referred to as \textit{average alloy} simulations.
These simulations mimic the effective average alloys, where each atom has the average properties of the alloy, similar to the approach used for Embedded Atom Method (EAM) type inter-atomic potentials \cite{Murray_1984}. An overview of the considered alloys and corresponding cluster sizes is given in Table\,\ref{tab:Alloys}

\begin{table}[h]
\centering
\caption{Overview of alloys, nanoscale cluster morphology and naming convention in the manuscript.}
\begin{tabular}{@{\extracolsep{4pt}}llll@{}}
\hline
Alloy & Nanoscale cluster configuration & Naming convention\\
\hline
CrMnFeCoNi & homogeneous equiatomic chemical composition &  average alloy\\
CrFeCoNiPd & homogeneous equiatomic chemical composition & average alloy\\
CrFeCoNiPd & 30$^3$\,nm$^3$ random solute clusters & configuration-I\\
CrFeCoNiPd & 15$^3$\,nm$^3$ random solute clusters & configuration-I\\
CoCu$_{1.71}$FeMnNi & 100\,nm lamellae & configuration-II\\
\hline
\end{tabular}
\label{tab:Alloys}
\end{table}

\section{Simulation Results}
\label{3_Res}
The material parameters for all MPEAs considered in the following theoretical calculations and simulations are given in Table \ref{tab:Material_parameters}.

\subsection{Analytical estimation of the CRSS of dislocation segment in MPEAs}
\label{sec:res_analytical_estimation}
As a first estimate of the influence of a uniform solute composition on the CRSS of a dislocation segment in MPEAs, Eqs. (\ref{eq:solute_stress})-(\ref{eq:deltaeb}) are evaluated for two of the MPEAs studied here: CrFeCoNiPd and CrMnFeCoNi, respectively. 

\begin{table}[h]
\centering
\caption{The material properties for the MPEAs considered in this study.}
\begin{tabular}{@{\extracolsep{4pt}}llll@{}}
\hline
Alloy & Shear Modulus $\mu$ \cite{ding_tuning_2019, okamoto2016size} & Poisson ratio $\nu$ \cite{okamoto2016size} & Burgers vector\\
& & & magnitude\\
\hline
CrMnFeCoNi & \SI{80}{GPa} &  $0.26$ & \SI{0.255}{\nm}\\
CoCu$_{1.71}$FeMnNi & \SI{80}{GPa} &  $0.26$ & \SI{0.255}{\nm}\\
CrFeCoNiPd & \SI{89}{GPa} & $0.26$ & \SI{0.255}{\nm}\\
\hline
\end{tabular}
\label{tab:Material_parameters}
\end{table}

\begin{figure}[h]
    \centering
    \includegraphics[width=0.5\textwidth]{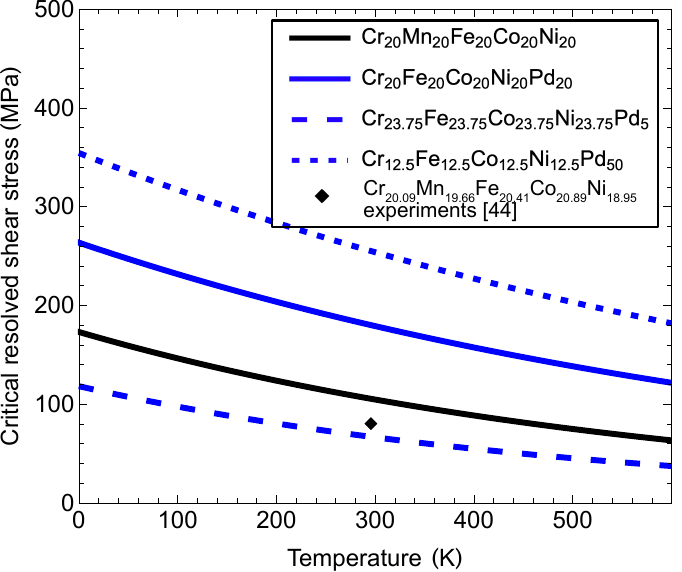}
    \caption{Critical resolved shear stress as a function of temperature as computed from Eq.\,(\ref{eq:solute_stress}) for the two MPEAs considered here at different atomic concentrations {and experimental estimate of the CRSS from a tensile deformation of single crystalline CrMnFeCoNi {\cite{kireeva2018twinning}}}. The effect of the temperature on the elastic parameters is neglected here for simplicity.}
    \label{fig:Solute_stress}
\end{figure}

The CRSS computed analytically for both CrFeCoNiPd and CrMnFeCoNi with an equal composition of all elements ($c_i = 0.2$) are shown in Fig. \ref{fig:Solute_stress} for the range of temperatures $0 \le T \le 600$K. Note, that for this calculation the temperature dependency of the elastic properties is ignored for simplicity. In the equi-atomic average alloy, the solute stress approximately doubles when replacing Mn with Pd at 300K. Additionally, two variations of the Pd concentration are shown: $c_\mathrm{Pd} = 0.05$ and $c_\mathrm{Pd} = 0.5$, respectively, where for all other elements $c_i = (1-c_\mathrm{Pd})/4$ with $i =$ Cr, Fe, Co, Ni. It is clear that varying the Pd concentration has a significant effect on the calculated CRSS with higher stresses obtained for larger Pd concentrations, in agreement with previous theoretical estimates  \cite{Yin2020b}. This effect of Pd can be attributed to its large atom radius (see Table\,\ref{tab:atomic_radii}), which results in a large atomic misfit as compared to the other elements in these two MPEAs. 
Experimental tensile tests of single crystalline CrMnFeCoNi at room temperature indicate a CRSS of $\approx82$\,MPa \cite{kireeva2018twinning}, as shown in Fig.\,\ref{fig:Solute_stress}. Those results are $\approx 18\,\%$ lower than the currently predicted CRSS at 300K, which is $\approx100$\,MPa. 
This difference can be rationalized by 1) differences in the initial dislocation density in the simulations versus experiments (the experiments do not report the initial dislocation density); or 2) an overestimation of the volume misfit of the various alloying elements in Eq.\,(\ref{eq:tauzero}) and Eq.\,(\ref{eq:deltaeb}) which is the key contribution in the model. For simplicity this volume misfit has been estimated here using atomic radii (Table\,\ref{tab:atomic_radii}),  which likely introduces some deviations from the real value. A more exact estimation of the volume misfit would have to account for further effects such as magnetism and could be determined by first principle simulations.
Still, the overall magnitude of the CRSS as well as the significant influence of Pd is captured well by the current model, considering that it is only based on elasticity constants and volume misfits and does not contain any adjustable parameters.

\subsection{Glide of an infinite long dislocation in MPEAs}
\label{sec:res_single_dislocation}

First, the glide of infinitely long dislocations at 300K through an effective average alloy with the average properties of CrFeCoNiPd (i.e. without any local variation in chemical composition and each atom has the average properties of a true random alloy comparable to the approach used for Embedded Atom Method (EAM) type inter-atomic potentials \cite{Murray_1984}); a field of $15^3$\,nm$^3$ random solute aggregation (Configuration-I) in CrFeCoNiPd; and a two-phase $\SI{100}{\nm}$ lamella system (Configuration-II) in CoCu$_{1.71}$FeMnNi are modeled. In all cases, the simulation volume is rectangular with edge lengths $5$ $\upmu$m parallel to the $[100]$, $[010]$, and $[001]$ directions, respectively. In the average alloy and Configuration-I simulations the dislocation belongs to the $[1\bar{1}0](111)$ slip system and is initially parallel to the $[11\bar{2}]$ direction. For Configuration-II simulations, we investigate a case where the dislocation is initially parallel to the $[11\bar{2}]$ direction (i.e. inclined to the lamella) and belongs to the $[1\bar{1}0](111)$ slip system as well as a case where the dislocation is initially parallel to the $[0\bar{1}1]$ direction (i.e. parallel to the lamella) and belongs to the $[0\bar{1}1](111)$ slip system. 

In all cases periodic boundary conditions are imposed on all directions. Thus, these simulations mimic the glide of an array of parallel infinite long dislocations. For the lamella system (Configuration-II) a constant uniaxial stress of $270$\,MPa in the range of the CRSS of the uniform alloy is imposed in the $[010]$ direction for the dislocation parallel to the lamella case and in the $[100]$ direction for the dislocation inclined to the lamella case. For the average alloy and Configuration-I simulations a constant uniaxial stress of $400$\,MPa in the range of the CRSS of the uniform alloy is imposed in the $[100]$ direction. 
The dislocation line in all simulations with infinitely long dislocations is discretized by a maximum segment size of 100$b$ (i.e. $\SI{25.5}{\nm}$) and minimum segment size of 25$b$ (i.e. $\SI{6.375}{\nm}$).
This numerical resolution ensures that the discretization of the dislocation line is well within the dimension of the considered nanoscale composition fluctuations in order to accurately capture local fluctuations of the dislocation motion.

\begin{figure}
  \subfloat[]{\scalebox{0.2}{\includegraphics{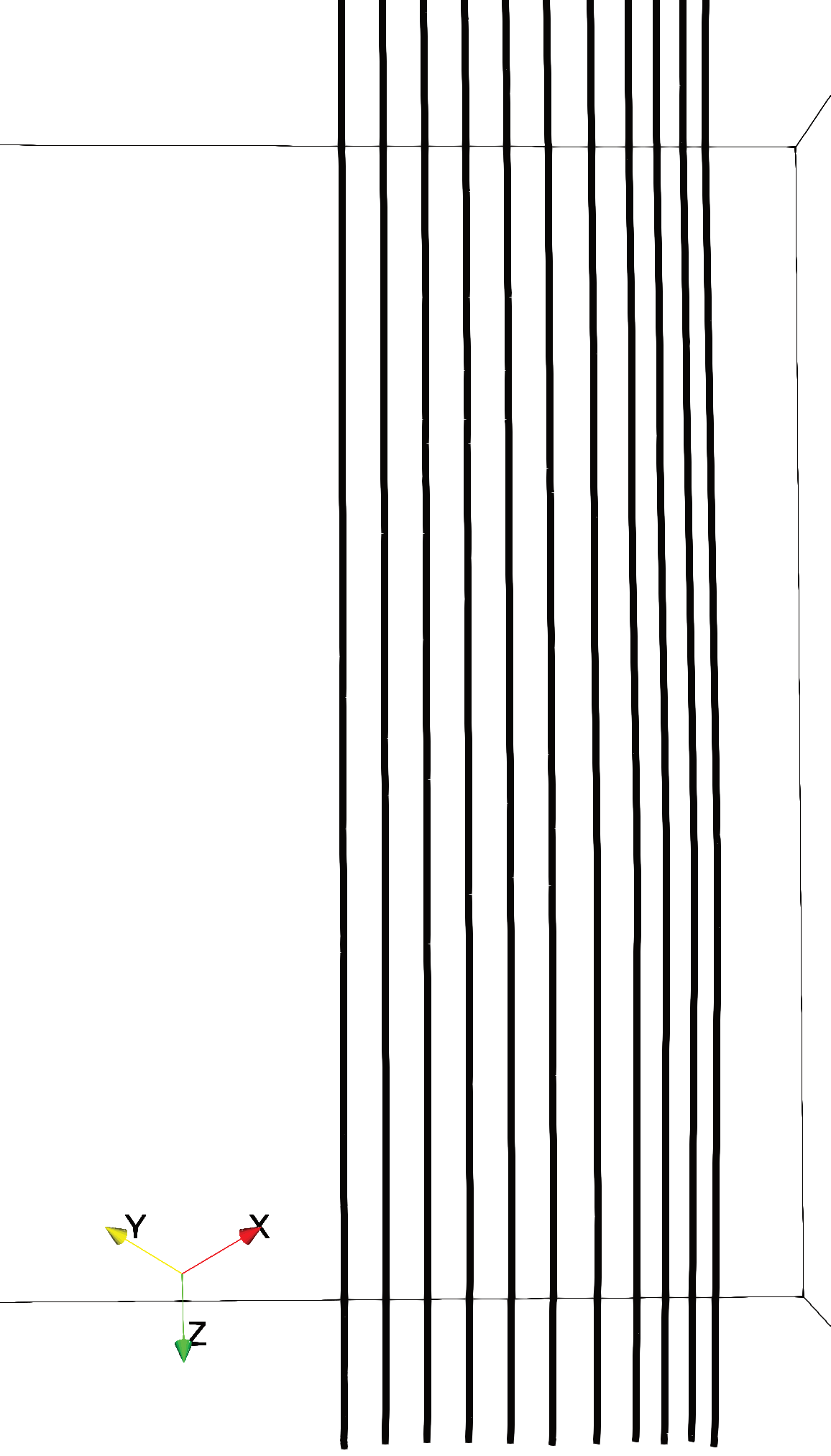}}\label{fig:Single_dislocation:a}}%
  \subfloat[]{\scalebox{0.18}{\includegraphics{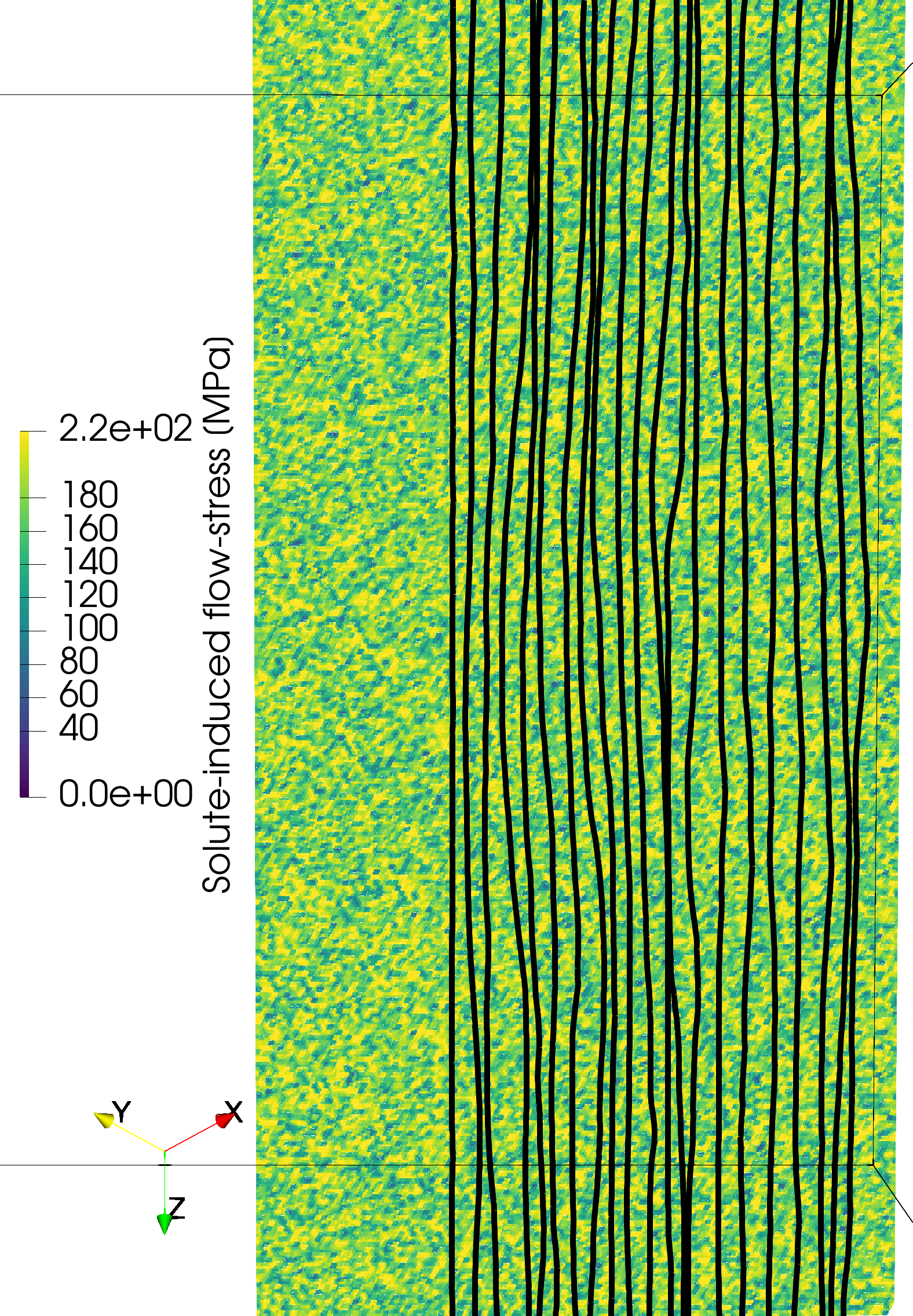}}\label{fig:Single_dislocation:b}}%
  \subfloat[]{\scalebox{0.18}{\includegraphics{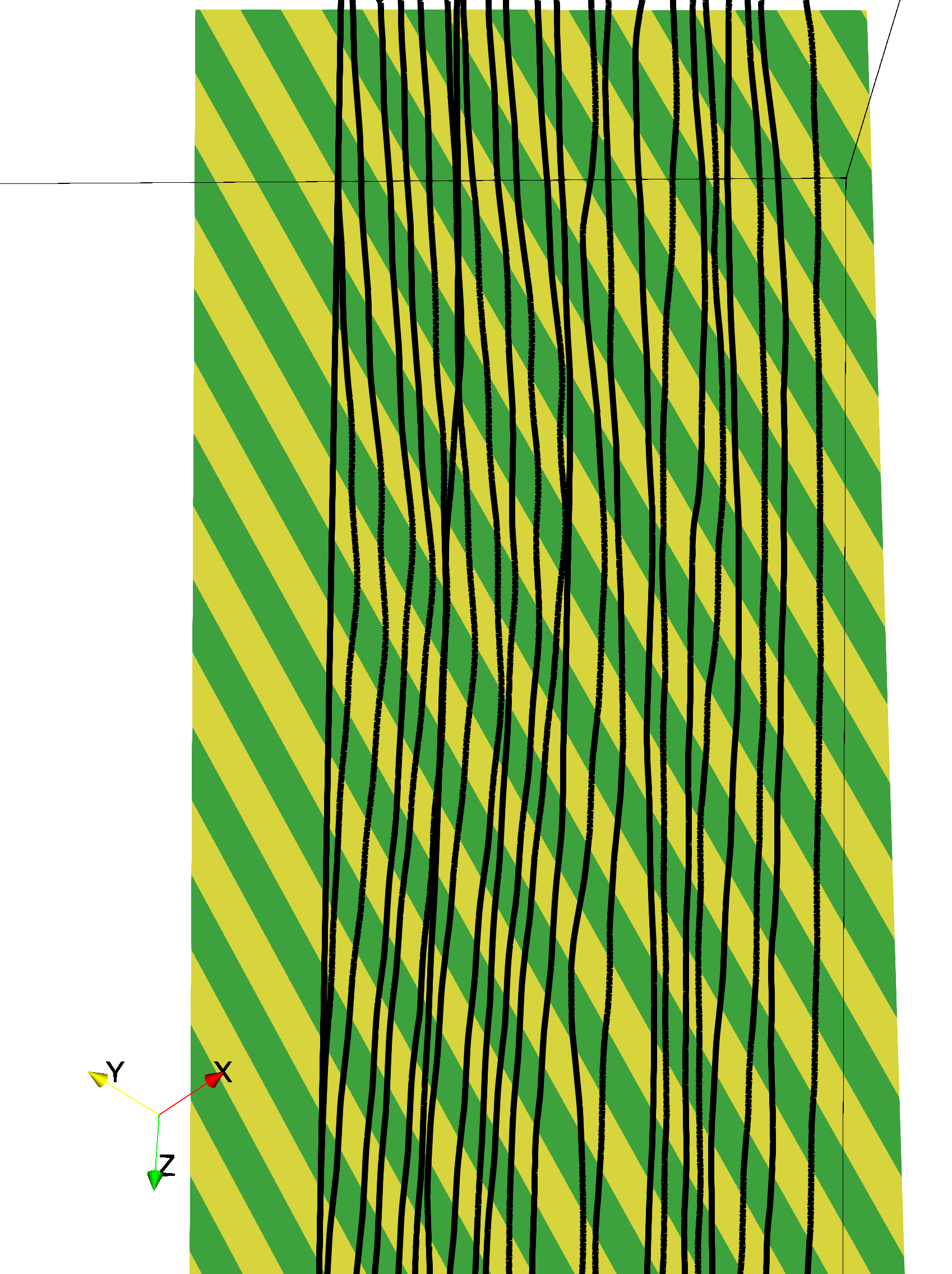}}\label{fig:Single_dislocation:c}}%
  \subfloat[]{\scalebox{0.049}{\includegraphics{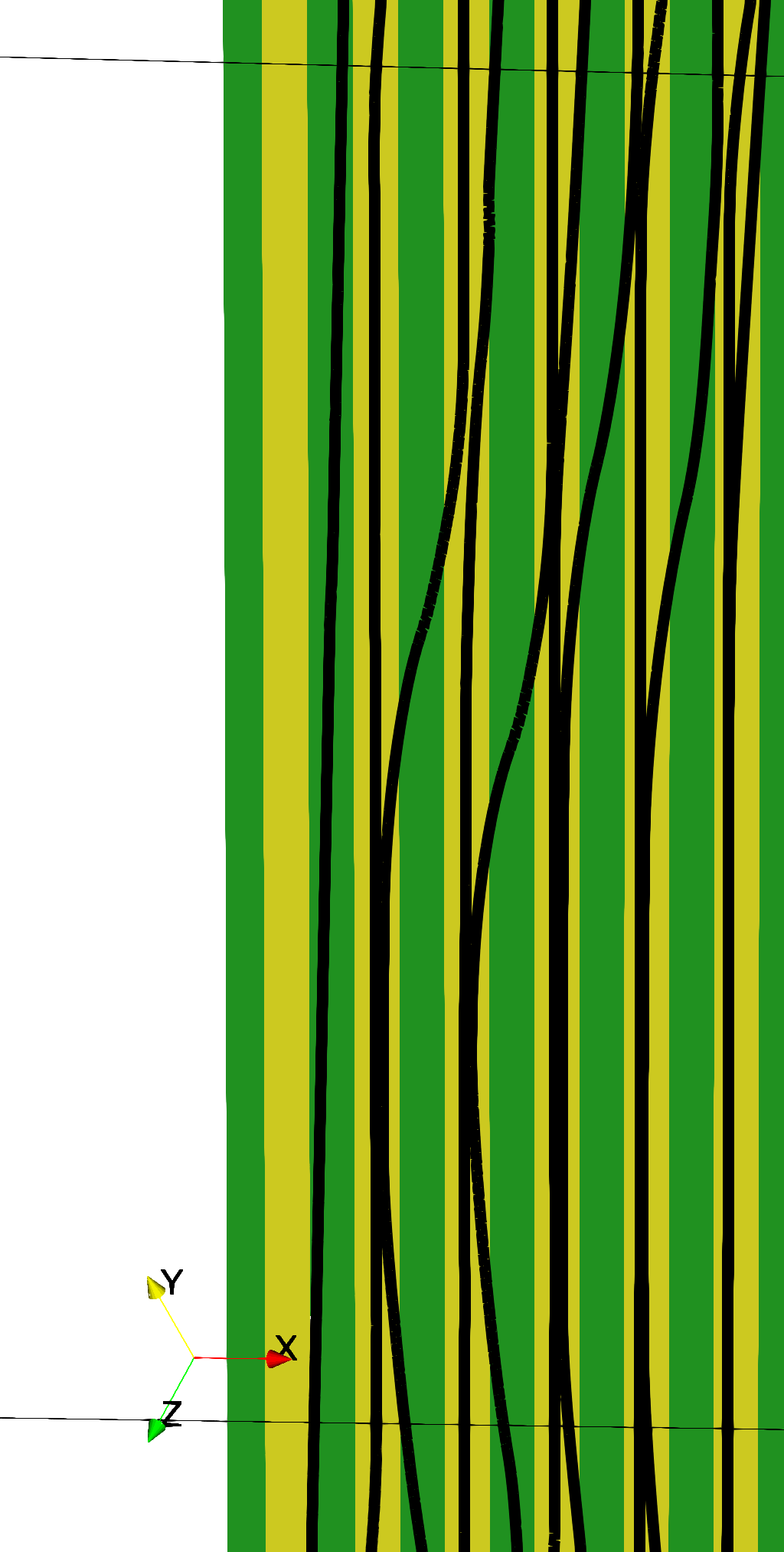}}\label{fig:Single_dislocation:d}}%
  \subfloat[]{\scalebox{0.37}{\input{Plastic_Strain.tex}}\label{fig:Single_dislocation:e}}%
  \caption{Superimposed snapshots of a single infinitely long edge dislocation moving through: (a) an effective average alloy with the average properties of CrFeCoNiPd; (b) a field of $15^3$\,nm$^3$ random solute aggregation (Configuration-I) in CrFeCoNiPd; and a two-phase $\SI{100}{\nm}$ lamella system in CoCu$_{1.71}$FeMnNi initially (c) inclined, and (d) perpendicular to the lamella. (e) The evolution of the plastic strain versus simulation time for all four cases. In all cases the simulation cell is periodic in all directions and the lamella in (c) and (d) is oriented parallel to the $[100]$ plane. The color-map in (b) show the distribution of the solute-induced resolved shear stress on the dislocation slip plane, while  in (c) and (d) show the lamella structure with the Cu rich phase in green and the Co-Fe rich phase in yellow.}\label{fig:Single_dislocation}
\end{figure}

Superimposed snapshots at different time steps showing the glide of the dislocation in the four different cases are shown in Fig. \ref{fig:Single_dislocation} (a)-(d), respectively. For the dislocation moving through the average alloy, a relatively continuous dislocation motion is observed without significant local changes of line orientation (see Fig.\,\ref{fig:Single_dislocation:a}). In contrast, for the random $15^3$\,nm$^3$ solute aggregation case, fluctuations along the dislocation line are observed due to the pinning and unpinning of different segments as the dislocation glides through the different local compositions (see Fig.\,\ref{fig:Single_dislocation:b}).
Thus, the dislocation glides in a consistent wavy manner in agreement with MD simulation observations (c.f. \cite{li2019strengthening, Antillon2020d, li2020core}).
A similar behavior can be observed in the two-phase 100 nm lamella system where the dislocation advances through the two lamella phases as shown in Fig.\,\ref{fig:Single_dislocation:c}.
Here, a similar wavy dislocation motion is observed as the dislocation is locally aligning with the interface of the lamellar solute aggregation features and creating a higher variation in the dislocation orientations.
An even stronger effect can be observed for a dislocation orientation parallel to the interface of the lamella phases as shown in Fig.\,\ref{fig:Single_dislocation:d}. 
In this case the dislocation is observed to exhibit a jump-like dislocation motion as alternating segments along the dislocation line advance through the lamella. 
The corresponding plastic strain evolution versus simulation time for all cases is also shown in Fig.\,\ref{fig:Single_dislocation:d}. 
A step-wise evolution can be clearly observed as the dislocation advances through the lamella structure in the parallel orientation case, in contrast to a more wavy evolution for the inclined to the lamella case. 
For the effective average alloy and the random $15^3$\,nm$^3$ solute aggregation cases, a more steady increase in plastic slip is observed.
The overall plastic strain rate for CoCu$_{1.71}$FeMnNi (Fig.\,\ref{fig:Single_dislocation:a} and b) is about $30\%$ higher than for CrFeCoNiPd (Fig.\,\ref{fig:Single_dislocation:c} and d) which can be simply rationalized by the different solute compositions in both alloys with a largely different dislocation CRSS (as shown in Fig.\,\ref{fig:Solute_stress}).

\captionsetup[subfigure]{position=top, singlelinecheck=false}

\subsection{Large scale simulations of dislocation plasticity in MPEAs with LCO}
\label{sec:res_bulk_system}

To study dislocation plasticity in MPEAs with LCO, large-scale 3D DDD simulations were conducted in cubical $5\times5\times5\,\upmu$m$^3$ simulation cells with periodic boundary conditions. The edges of the cubical cell are aligned with the $\langle100\rangle$ crystal directions. A uniaxial tensile loading is imposed along the $[100]$ crystal direction at a strain rate of $\dot{\varepsilon}=200$\,s$^{-1}$ and the temperature is fixed at \SI{300}{K}. The initial dislocation configuration consists of randomly positioned Frank-Read sources having random orientation and random lengths in the range between 500\,nm and 2000\,nm. The initial dislocation density is $\rho \approx 3.0\times 10^{12}~$m$^{-2}$ and the elastic parameters used in the simulations are given in Table \ref{tab:Material_parameters}. The simulations were conducted for at least two different statistically equivalent random initial seeds of the initial dislocation microstructure for each alloy, and no relevant difference in overall stress-strain behavior was observed between the different seeds, respectively. All dislocation lines in these simulations are discretized using a maximum segment size of 400$b$ ($\SI{102}{\nm}$) and minimum segment size of 100$b$ ($\SI{25.5}{\nm}$). 
The resolution of the dislocation line is dynamically adapted within the range prescribed by the maximum and minimum segment size to account for location dislocation curvature.
Thus, the numerical resolution of the dislocation line is chosen to balance the computational resources and the requirements for a small enough dislocation segment size to adequately represent the effects of the composition fluctuations on the dislocation motion.

\begin{figure}[h]\centering
\begin{minipage}{0.41\textwidth}\centering
\subfloat[]{\scalebox{0.5}{\input{Stress_Strain_comparison_tex.tex}}\label{fig:Stress_strain_Density_periodic:a}}\\%
\vspace{-2em}
\subfloat[]{\scalebox{0.5}{\input{Total_Density_comparison_tex.tex}}\label{fig:Stress_strain_Density_periodic:b}}%
\end{minipage}
  \begin{minipage}{0.49\textwidth}\centering
  \subfloat[]{\scalebox{0.6}{\includegraphics{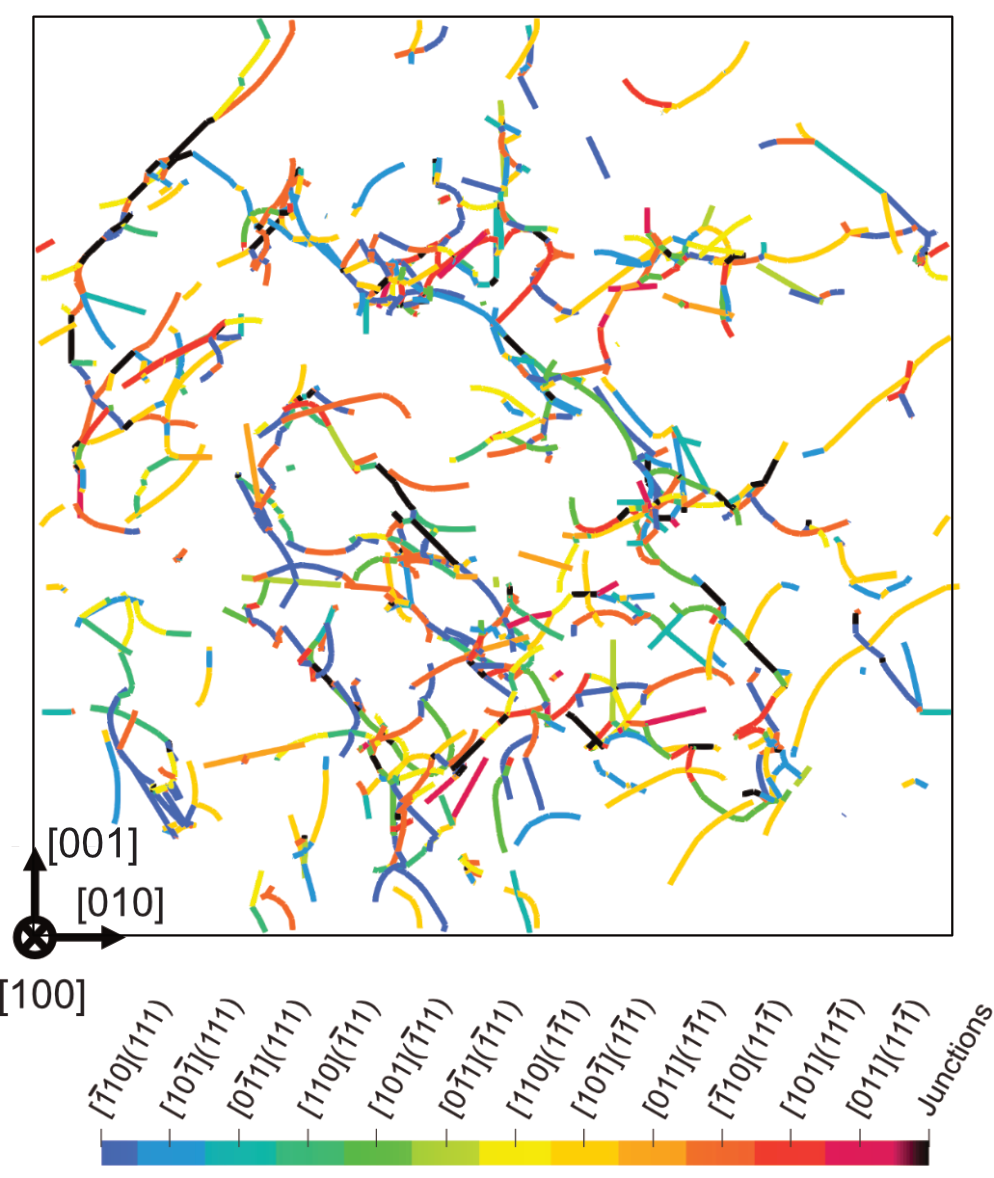}}\label{fig:Stress_strain_Density_periodic:c}}%
  \end{minipage}
  \caption{(a) Engineering stress and (b) total dislocation density versus engineering strain as predicted from the current 3D DDD simulations for: CrFeCoNiPd and CrMnFeCoNi modeled as effective average alloys without any LCO, CrFeCoNiPd with 15$^3$\,nm$^3$ random solute clusters, CrFeCoNiPd with 30$^3$\,nm$^3$ random solute clusters, and CoCu$_{1.71}$FeMnNi with 100 nm lamellae parallel to the $[100]$ plane. 
  (c) The predicted dislocation microstructure in a thin slab having thickness 400\,nm normal to the loading direction ($X$-axis) that was extracted from the center of the 3D DDD simulation volume for the CrFeCoNiPd alloy with 15$^3$\,nm$^3$ random solute clusters at 0.35\% strain. The different line colors in (c) indicate the dislocation slip system as shown in the associated color-map.}\label{fig:Stress_strain_Density_periodic}
\end{figure}

The stress-strain curves for CrFeCoNiPd and CrMnFeCoNi modeled as effective average alloys without any local chemical composition variation, CoCu$_{1.71}$FeMnNi alloy with 100 nm lamellae, CrFeCoNiPd alloy with 15$^3$\,nm$^3$, and CrFeCoNiPd alloy with 30$^3$\,nm$^3$ random solute clusters as predicted from the current 3D DDD simulations are shown in Fig.\,\ref{fig:Stress_strain_Density_periodic:a}. 

\captionsetup[subfigure]{position=top, singlelinecheck=false}

Comparing the stress-strain curves for CrFeCoNiPd and CrMnFeCoNi modeled as effective average alloys it is observed that the yield stress almost doubles by exchanging Mn with Pd. 
Additionally, the CrFeCoNiPd simulations with two random solute cluster sizes show consistently higher stresses compared to their uniform counterparts. However, no significant difference between the two solute cluster sizes is observed. On the other hand, the flow stress predicted at $0.16\%$ strain from the 3D DDD simulations of CoCu$_{1.71}$FeMnNi with 100 nm lamellae is $\approx$300\,MPa (i.e. resolved shear stress of $\approx$123\,MPa with Schmid-factor of 0.41). This value is in good agreement with the CRSS of the lamella layer having the higher CRSS (\SI{118}{MPa}), indicating that this harder lamella controls the strength of this alloy.

The evolution of the dislocation density versus strain for the two uniform average alloy simulations (no LCO) as well as that for the CoCu$_{1.71}$FeMnNi alloy with 100 nm lamellae are qualitatively similar as shown in Fig.\,\ref{fig:Stress_strain_Density_periodic:b}.
On the other hand, the dislocation density increase in the two cases of random LCO show a significantly higher rate with the dislocation density reaching twice the dislocation density for the uniform average alloy simulations case at $0.4\,\%$ total strain.
The dislocation structure in a 400 nm thick TEM-like foil normal to the loading direction extracted from the center of the simulation volume of the random $15^3$\,nm$^3$ LCO case at 0.35\% strain is shown in Fig.\,\ref{fig:Stress_strain_Density_periodic:c}.
The dislocation structure is characterized by a complex entanglement of dislocations on different slip systems (shown by the different colors) driven by an abundance of cross-slipped dislocation segments and dislocation junctions (shown as black segments).

\begin{figure}[htbp]\centering
  \subfloat[]{\scalebox{0.42}{\input{BulkCrossSlip_comparison_plstrain.tex}}\label{fig:Crossslip_periodic:a}}%
  \subfloat[]{\scalebox{0.31}{\includegraphics{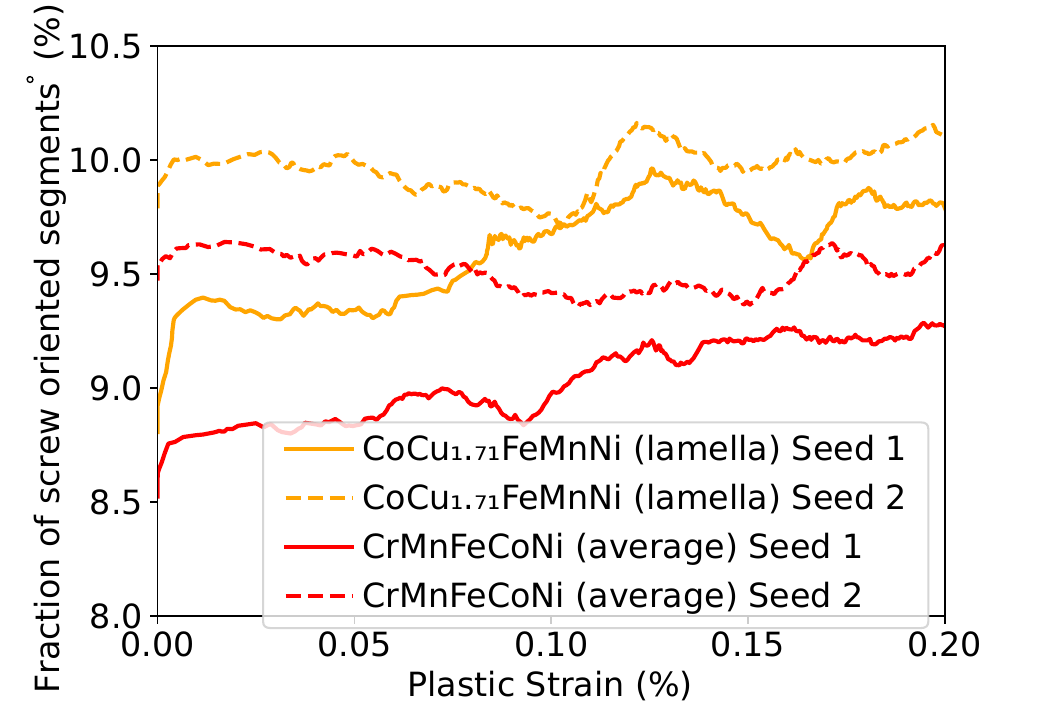}}\label{fig:Crossslip_periodic:b}}%
  \caption{{The number of bulk-type cross-slip events versus plastic strain for each simulation shown in Figure} \ref{fig:Stress_strain_Density_periodic} (b), and Fraction of the dislocation density that is screw oriented (segments oriented within $\pm$15\% of the Burgers vector direction) for the CoCu$_{1.71}$FeMnNi alloy with 100 nm lamellae compared to the average alloy simulations (c). In (c), results from simulations with two random seeds of initial dislocation distributions are shown.}\label{fig:Crossslip_periodic}
\end{figure}

To explain the strong dislocation density increase for the simulations with the random LCO alloys, the cross-slip activity as a function of plastic strain is analyzed and shown in Fig.\,\ref{fig:Crossslip_periodic:a}. 
{Here, the accumulated number of bulk-type cross-slip events are shown for the different simulation cases. 
We note that the restriction to bulk-type cross-slip is intended to isolate the solute-induced effect on the cross-slip activity, since this cross-slip type is not influenced by intersecting dislocations as is the case for intersection-type cross-slip.}
{For CrFeCoNiPd with random solute clusters the number of bulk-type cross-slip events in both cluster sizes is almost two orders of magnitude higher as compared to CrFeCoNiPd and CrMnFeCoNi modeled as effective average alloys.}
It is also interesting to note, that both average alloys show similar total cross-slip activity despite the difference in the stress levels in both alloys.
This indicates a significant influence of bulk cross-slip activation energy fluctuations in the random LCO simulations.

\captionsetup[subfigure]{position=top, singlelinecheck=false}

Furthermore, in the CoCu$_{1.71}$FeMnNi lamella-type LCO simulations, the number of bulk-type cross-slip events at the same plastic strain is about $100\,\%$ higher compared to the average alloy case. 
This is noticeable and worth investigating more closely since for the lamella LCO alloy the cross-slip activation energy is the same throughout the simulation, as explained in Section \ref{sec:solute_aggregation}.
The fraction of the dislocation density that is screw oriented (defined here as the fraction of dislocation segments that are oriented $\pm 15^\circ$ from the dislocation Burgers vector) for the CoCu$_{1.71}$FeMnNi lamella LCO simulations compared to the average alloy simulations for two different statistically equivalent random initial seeds of the initial dislocation microstructure are shown in Fig.\,\ref{fig:Crossslip_periodic:b}.
For both seeds, a higher fraction of screw orientations are observed in the lamella LCO compared to the uniform alloy throughout the simulation, indicating an effect of the lamella on increasing the probability of dislocation segments aligning in the screw orientation.
This can be rationalized by the fact that the lamella is oriented parallel to the $[100]$-plane, making it likely that dislocation segments are reoriented in the screw orientation due to pinning and unpinning effects and thus have a higher cross-slip probability.

\section{Discussion}
\label{4_Disc}
Here, a model for solute-dislocation interactions was introduced in the framework of three-dimensional discrete dislocation dynamics simulations by accounting for the local resistance of solute induced atomic misfit on dislocation glide, known as solute strengthening, and the solute-induced effects on cross-slip.
Analyzing the motion of single dislocations through a field of random solute aggregation clusters (Fig.\,\ref{fig:Single_dislocation:b}), or lamella-like solute aggregation (Fig.\,\ref{fig:Single_dislocation:c} and \ref{fig:Single_dislocation:d}), it can be concluded that local changes in solute composition leads to a wavy dislocation motion in the alloy incorporating random LCO, which is characterized by pinning and unpinning effects of the dislocations at solute clusters (Fig.\,\ref{fig:Single_dislocation:b}). These observations are in good agreement with MD observations of dislocation motion through MPEAs with LCO (e.g. \cite{li2019strengthening, Antillon2020d}).

In the case of lamella-like LCOs, when the dislocation line is oriented parallel to the lamella (Fig.\,\ref{fig:Single_dislocation:c}) the pinning and unpinning effects lead to the formation of dislocation alignments along the interfaces with short segments spanning in between during the dislocation motion.
This observation is in agreement with experimentally observed microstructural alignments during deformation of homogenized CoCu$_{1.71}$FeMnNi \cite{shim_nanoscale_2019}.
It should be noted that in order to show such microstructure formation in large scale DDD bulk simulations, significant computational resources would be needed to accommodate for high dislocation densities alongside with the requirement of fine dislocation discretization.
The latter requirement can be rationalized by considering that the numerical resolution of the dislocation line (i.e. the dislocation segment size) has to be adjusted to the size of the chemical resolution in order to accurately capture the local dislocation curvature induced by pinning of the dislocation at sites of high resistive stress.

Similarly, the size of the random solute aggregation clusters (Configuration-I) chosen in the current 3D DDD simulations are larger than those reported in MD simulations \cite{li2019strengthening, Antillon2020d} due to numerical constraints. This results in larger distances between pinning sites compared to MD simulations, thus a larger dislocation wavelength. 
Thus, the corresponding wavelength depends on the relation between the numerical resolution and the cluster size. 
Smaller wavelengths will be obtained in DDD with even lower cluster sizes and higher numerical resolution. 
However, it will not be reasonable to attempt DDD simulations using such high numerical resolution since this would be attempting to model atomic length scales, which is beyond the scope of the corse-grained DDD technique.
Nevertheless, the same wavy dislocation motion observed in MD simulations and experiments are also observed in the current DDD simulations.
Furthermore, this wavy motion of the dislocation in the current 3D DDD simulations leads to a higher variation in the dislocation orientations along its length (Fig.\,\ref{fig:Single_dislocation:b}), which can increase the probability of finding dislocation segments aligned in a screw orientation.
The same conclusion holds for the motion of the dislocation through the lamella-type LCO (Fig.\,\ref{fig:Single_dislocation:c} and \ref{fig:Single_dislocation:d}).

The interaction of dislocations with local random solute aggregation clusters or lamella-like solute aggregation has also a strong effect on the response of dislocation ensembles. 
The flow strength at 0.25\% strain for CrMnFeCoNi modeled as effective average MPEA as predicted from the current 3D DDD simulations mimicking a single crystal under uniaxial tension loading along the $[100]$ direction is $\approx 260~$MPa, while that for CrFeCoNiPd modeled as effective average MPEA is $\approx 440$ MPa. 
Experimental tensile tests on $[100]$-oriented single crystalline CrMnFeCoNi at room temperature showed a yield stress of $\approx 200~$MPa \cite{kireeva2018twinning}, which is lower than the corresponding flow strength predicted by the 3D DDD simulations.
A more exact value could be obtained by a more accurate estimate of the misfit volumes of the constituting elements with first principle simulations and by using an experimentally measured initial dislocation density (which was not reported in literature \cite{kireeva2018twinning}).
It is also interesting to note that the overall magnitude and the difference in flow strength if converted into a uniaxial flow strength for an fcc polycrystal by multiplying by the Schmid-factor ($\approx 0.41$) and the Taylor-factor (3.06) agree fairly well with the experimentally reported flow strength for polycrystalline CrMnFeCoNi and CrFeCoNiPd multi-component alloys \cite{ding_tuning_2019, shim_nanoscale_2019}. 
The single crystal flow strength at 0.25\% strain of CrFeCoNiPd with $15^3$\,nm$^3$ random LCO as predicted from the 3D DDD simulations is $\approx 460~$MPa.
Comparing this value to the flow strength predicted for CrFeCoNiPd modeled as effective average MPEA ($\approx 440$ MPa) indicates a rather secondary effect of the fluctuations in the local solute composition on strength as compared to the pure Pd misfit effect (Fig.\,\ref{fig:Stress_strain_Density_periodic:a}).
Thus, the 3D DDD simulations show that the large increase in flow stress observed in experiments \cite{ding_tuning_2019} by replacing Mn with Pd in the CrMnFeCoNi MPEA can be explained entirely by the higher misfit of Pd atoms rather than nanoscale composition fluctuation.
In addition to using a linear pure Ni mobility law, these conclusions hold also for the response of the CrFeCoNiPd effective average alloy and the 100\,nm lamella LCO simulation using a linear concentration-dependent mobility law for FeNiCr-based stainless steels recently developed using atomistic simulations in \cite{Chu2020} (see \ref{sec:appendix}).

The solute strengthening formulation used here relies on an assumption of a full randomness attributed to the size misfit between atoms. 
Furthermore, the smallest solute clusters investigated in the current study are still an order of magnitude larger than some experimentally and computationally reported cluster sizes, which are on the order of 1-3 nm \cite{li2019strengthening, ding_tuning_2019}.
Both assumptions neglect an additional hardening component that can be attributed to solute-solute interaction effects when a moving dislocation shifts atomic layers with present chemical ordering on the range of nearest atomic neighbors \cite{Antillon2020d}.
Only recently an extension to the Labusch-Varvenne solute strengthening model has been proposed by Antillon et al. \cite{Antillon2020d} assuming an additional correlation between atoms in different atomic rows. 
Furthermore, a microelasticity approach to obtain the CRSS of MPEAs could be used to introduce the effect of spacial correlations of dislocation core effects into 3D DDD  \cite{geslin2021microelasticity,geslin2021microelasticitya}.
It might be expected that by incorporating these additional effects, an even stronger hardening response would be induced by solute composition fluctuations. 

The major consequence of the solute-dislocation interactions on the overall dislocation activity is the increase in bulk-type cross-slip activity as compared to the reference simulations incorporating a uniform solute distribution throughout the simulation volume (i.e. effective average MPEA).
This can be seen in Fig.\,\ref{fig:Crossslip_periodic:a}.
The increase in cross-slip activity can be attributed to two different aspects:
(i) a larger variation of dislocation orientations induced by the wavy and sluggish dislocation motion caused by locally high pinning stresses due to a locally varying solute compositions; and
(ii) fluctuation in the cross-slip activation energy in the random LCO alloy, resulting in a significant increase in cross-slip activity compared to the effective average alloy.
Both aspects are consistent with experimental observations of intense cross-slip in some MPEAs \cite{ding_tuning_2019, ding2019real}, as well as observations in MD simulations of random alloys showing enhanced dislocation cross-slip activity \cite{nohring_dislocation_2017, nohring_cross-slip_2018}. 
It should be noted that in the current analysis the focus was on the bulk-type cross-slip mechanism as defined in \cite{hussein_microstructurally_2015} since bulk cross-slip does not depend on additional requirements such as the dislocation intersecting the free surface or intersecting with forest dislocations.
Thus, the effect of the CRSS induced by chemical composition on cross-slip will primarily be noticeable in bulk-type cross-slip.
Furthermore, bulk-type cross-slip accordingly represents the same cross-slip mechanism studied in MD simulations of random alloys \cite{nohring_dislocation_2017}.
Comparing the influence of the two aspects mentioned above, it is obvious from Fig.\,\ref{fig:Crossslip_periodic:a} that the effect of fluctuations in the activation energy on the bulk-type cross-slip activity is significantly higher than the effect caused by locally high pinning stresses.
It is expected, however, that the latter effect will become more significant, when the dislocation segment size is chosen as small as the size of the solute clusters in CrFeCoNiPd or the lamella spacing in CoCu$_{1.71}$FeMnNi, because a higher variation in dislocation line orientations will be induced then by facilitating the creation of local dislocation curvature. 
This in return increases cross-slip activity and mutual dislocation interaction.
It should be pointed put that more accurate and chemical composition-dependent cross-slip parameters are likely to allow for a more accurate representation of the cross-slip process in MPEAs. This is especially the case for the two-phase CoCu$_{1.71}$FeMnNi alloy which incorporates a significantly different chemical composition in both phases.

While the current results indicate that the ad-hoc approach of randomly choosing the bulk cross-slip activation energy can reproduce the experimental observations of intense cross-slip in some MPEAs, it is still not a fully satisfying or generalizable approach, because in other experimental observations planar slip was also reported \cite{Zhang2020}.
This suggests the need for further in-depth analysis of the cross-slip in multi-component alloys relating relevant parameters such as the stacking fault energy or the Escaig stress difference between the primary and the cross-slip plane to local solute composition fluctuations.
Through such analysis a more physical based cross-slip model for 3D DDD incorporating a dependency on local chemical composition could be derived.
Despite this open question, it can be concluded from the present results that fluctuations in local chemical solute composition as well as in cross-slip activation energy significantly increases cross-slip activity leading to increased dislocation multiplication and the formation of dense dislocation networks, as shown for the random LCO alloy in Fig.\,\ref{fig:Stress_strain_Density_periodic:c}.
This is in agreement with experimental microstructures for the investigated alloys in \cite{ding_tuning_2019}.

\section{Summary and Conclusion}
\label{5_Conc}

An approach to model the interaction of dislocations with substitutional solutes in fcc alloys in 3D DDD simulations was developed. In this approach the influence of the solute-dislocation interaction on dislocation glide and bulk-type cross-slip were accounted for.
Using this model, it was shown that by incorporating local fluctuations in the chemical composition in 3D DDD simulations, key experimental observations regarding the mechanical behavior and dislocation activity in CrCoFeNi-type MPEAs can be explained.
The current 3D DDD simulation results show that the wavy and sluggish dislocation motion is induced by pinning effects from high local flow-stress due to atomic misfit.
Such discontinuous dislocation motion leads to strong variations in dislocation orientations increasing the probability of dislocation segments aligning with the screw orientation.
Additionally, the fluctuations in the cross-slip activation energy leads to substantial cross-slip activity resulting in a high dislocation density increase in agreement with experimental observations.
Furthermore, dislocation alignment with lamellar solute aggregation features are observed on the level of individual dislocations, providing an insight into the mechanisms causing the formation of experimentally observed microstructures in CoCu$_{1.71}$FeMnNi.
{Our simulation results demonstrate that the experimental behavior cannot be reproduced by assuming a perfect solid solution. 
Instead, the incorporation of chemical variations and their effect on the CRSS and cross-slip activity is necessary to accurately predict the dislocation activity and mechanical behavior of MPEAs.}

More substantially, the current implementation of solute dislocation interactions in 3D DDD can also be applied to any fcc alloy containing substitutional solutes and possible local fluctuations thereof, such as in additive manufacturing of stainless steels or nickel-based alloys \cite{wang_additively_2018,liu_dislocation_2018, ZHANG2018200}.
The presented approach therefore provides a promising step towards modeling micromechanisms and dislocation microstructure evolution in real alloys.
Nevertheless, several modifications can also be incorporated to the current model to more accurately model solute-induced effects on dislocation activity. 
First, the effect of chemical short-range order on strain hardening could be more accurately reproduced by accounting for a correlation of neighboring solutes for the interaction of the solute with the dislocation as recently proposed \cite{Antillon2020d,geslin2021microelasticity,geslin2021microelasticitya}.
Second, an improved cross-slip model could be formulated to account for the influence of local chemical composition on cross-slip parameters such as the activation energy of different cross-slip types and on the Escaig stress.
Finally, by including the effect of interstitial atoms on dislocation glide (e.g. \cite{gu_theoretical_2020}) as well as atomic diffusion, a more complete description of the dislocation activity and the mechanical behavior in real fcc alloys could be provided.

\section*{Acknowledgements}

This work was supported by the U.S. National Science Foundation CAREER award CMMI-1454072 and DMR award DMR-1807708 and by the Office Of Naval Research through grant number N00014-18-1-2858.
The authors further acknowledge computational resources at the Maryland Advanced Research Computing Center (MARCC).

\appendix

\section{Mechanical response of MPEAs using different dislocation mobility laws}
\label{sec:appendix}

\begin{figure}[h!]
    \centering
    \scalebox{0.5}{\input{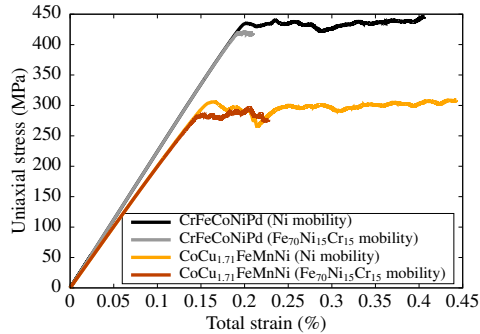}}
    \caption{Effect of using a pure Ni mobility law versus a Fe$_{70}$Ni$_{15}$Cr$_{15}$ mobility law \cite{Chu2020} on the stress-strain response of CrFeCoNiPd effective average alloy (No LCO) and CoCu$_{1.71}$FeMnNi 100nm Lamella LCO.}
    \label{fig:Stress_strain_FeCrNi}
\end{figure}

In Section \ref{sec:res_bulk_system}, the mechanical response in bulk DDD simulations is investigated using a dislocation mobility law for pure Ni. 
Recently, a temperature and composition-dependent dislocation mobility law for Fe$_{70}$Ni$_{x}$Cr$_{0.3-x}$-based stainless steels has been developed using MD simulations \cite{Chu2020}.
The authors fitted atomistic mobility data for screw dislocations using a Ni composition between 0.0 and 30.0 at.\% to a linear mobility law and determined a linear solute drag coefficient.
While the dislocation mobility of FeNiCr-based stainless steels does not fully represent the dislocation mobility of CrFeCoNi-based MPEAs, it incorporates more relevant physics in contrast to pure Ni. 
Fig.\,\ref{fig:Stress_strain_FeCrNi} compares the stress-strain curves using the Fe$_{70}$Ni$_{15}$Cr$_{15}$ mobility law in the effective average alloy model of CrFeCoNiPd as well as CoCu$_{1.71}$FeMnNi 100nm lamella LCO with that using a Ni dislocation mobility law.
It is observed that the flow stresses predicted by both mobility laws are in the same range, indicating that the influence of the dislocation mobility law on the overall yield behavior is negligible.

% \section*{References} 
\bibliographystyle{unsrt}

% \bibliography{Literatur_markus}

% \end{thebibliography}
\end{document}